\documentclass[%
 reprint,
nobibnotes,
 amsmath,amssymb,
 aps,
floatfix,
]{revtex4-2}

\usepackage{graphicx}
\usepackage{dcolumn}
\usepackage{bm}
\usepackage{mathtools}
\usepackage{braket}
\usepackage[caption=false, position=top]{subfig}
\usepackage{lipsum}
\usepackage{hyperref}

\newcolumntype{P}[1]{>{\centering\arraybackslash}p{#1}}
\newcolumntype{M}[1]{>{\centering\arraybackslash}m{#1}}

\newcommand{\eps}{\varepsilon}
\renewcommand{\vec}[1]{\boldsymbol{\mathrm{#1}}}
\renewcommand{\exp}[1]{\text{e}^{#1}}

\newcommand{\op}[1]{\hat{#1}}
\newcommand{\vop}[1]{\hat{\vec{#1}}}

\newcommand{\unvec}[1]{\hat{\vec{#1}}}
\newcommand{\abs}[1]{|#1|}

\newcommand{\norm}[1]{|{#1}|}

\begin{document}

\preprint{APS/123-QED}

\title{Exciton absorption, band structure, and optical emission in biased bilayer graphene}

\author{Mikkel Ohm Sauer$^{1,2,3}$}
\email{mikkelos@mp.aau.dk}
\author{Thomas Garm Pedersen$^{1,3}$}
\email{tgp@mp.aau.dk}
\affiliation{$^1$Department of Materials and Production, Aalborg University, 9220 Aalborg {\O}st, Denmark}
\affiliation{$^2$Department of Mathematical Sciences, Aalborg University, 9220 Aalborg {\O}st, Denmark}
\affiliation{$^3$Center for Nanostructured Graphene (CNG), 9220 Aalborg {\O}st, Denmark}

\date{\today}

\begin{abstract}
Biased bilayer graphene (BBG) is a variable band gap semiconductor, with a strongly field-dependent band gap of up to $300 \, \text{meV}$, making it of particular interest for graphene-based nanoelectronic and -photonic devices. The optical properties of BBG are dominated by strongly bound excitons.
We perform \textit{ab initio} density-functional-theory+Bethe-Salpeter-equation modelling of excitons in BBG and calculate the exciton band structures and optical matrix elements for field strengths in the range $30-300 \, \text{mV/Å}$. 
The exciton properties prove to have a strong field dependence, with both energy ordering and dipole alignment varying significantly between the low and high field regions. 
Namely, at low fields we find a mostly dark ground state exciton, as opposed to high fields, where the lowest exciton is bright. 
Also, excitons preferentially align with a dipole moment opposite the field, due to the field-induced charge transfer in the ground state of BBG. However, in stronger fields, this alignment becomes energetically less favorable. 
Additionally, the bright excitons show particle- and light-like bands similar to monolayer transition metal dichalchogenides.
Finally, we model the radiative lifetimes and emission properties of BBG, which prove to be strongly dependent on temperature in addition to field strength.
\end{abstract}

\maketitle

\section{Introduction}
Since the successful exfoliation of graphene in 2004 \cite{article:grapheneOG}, intense research into its unique optical, electronic, and structural properties has unveiled fascinating phenomena such as an anomalous quantum Hall effect \cite{article:Novoselov_2005, article:Zhang_2005, article:Novoselov_2006} and massless Dirac fermion physics \cite{article:Geim1,article:Katsnelson_2006,article:Huard1}. By adding an additional layer, bilayer graphene (BG) with entirely different, yet equally fascinating, electronic properties is produced. While both pristine monolayer and bilayer graphene are conductors, a particularly striking property of bilayer graphene is the possibility of introducing a non-zero band gap. This is typically achieved by applying a gate voltage in the out-of-plane direction, and thus obtaining biased bilayer graphene (BBG) \cite{article:BBG1,article:BBG3,article:BBG4,article:BBG8,article:BBG9,article:BBG10,article:BBG12}. The bias breaks the sublattice symmetry and induces a continuously voltage-tunable band gap varying from mid- to far-infrared \cite{article:BBG1,article:BBG3,article:BBG4,article:BBG8,article:BBG9,article:BBG10,article:BBG12}, relevant for a multitude of scientific and technological applications. Furthermore, the broken symmetry leads to dipole-allowed even-order optical nonlinearities \cite{article:BBG5}.

\begin{figure*}[htp]
    \centering
    \includegraphics[width=.8\linewidth]{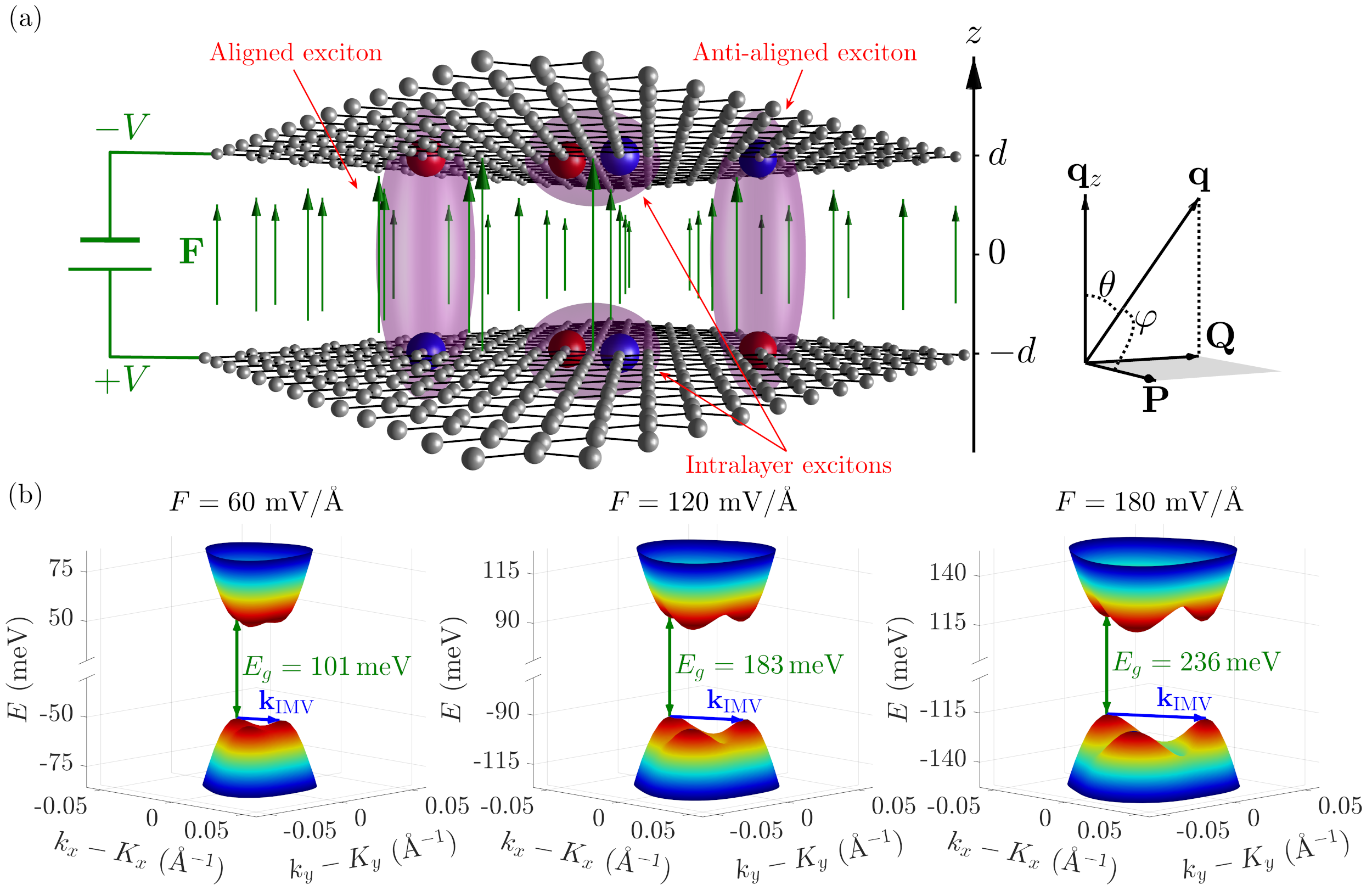}
    \caption{(a) Schematic structure of bilayer graphene in an external electric field $\vec{F}$. The ellipsoids represent four different types of excitons found in BBG, namely aligned and anti-aligned interlayer excitons, and top and bottom intralayer excitons. Here, holes and electrons are represented by red and blue spheres, respectively. The vector diagram on the right shows center of mass momentum $\vec{Q}$, transition matrix element $\vec{P}$, and photon momentum $\vec{q}$. (b) Surface plots of the calculated DFT band structures near the Dirac-point $K$, including the band gap and inter-mini-valley vector $\vec{k}_{\text{IMV}}$. Note that the $k_x$- and $k_y$-scales are equal for the different external fields, illustrating the increasing inter-mini-valley distance. Furthermore, the energy scale is the same in all plots, but with different discontinuous jumps around $E = 0$.}
    \label{fig:introduction}
\end{figure*}

The optical response of two-dimensional semiconductors is characterized by intense excitonic effects due to quantum confinement and significantly lower screening of the Coulomb potential compared to bulk materials. Indeed, Bethe-Salpeter equation (BSE) calculations for BBG show a noticeable reduction of the optical band gap due to excitons \cite{article:BBG1,article:BBG3}. Furthermore, the formation of an excitonic insulator phase has been observed through tuning of the bias and substrate \cite{article:BBG13}. Additionally, in a weak bias, BBG exhibits an almost entirely dark singlet excitonic ground state due to its unique pseudospin texture \cite{article:BBG1,article:BBG3}. In the present work, we demonstrate that the lowest bright excitons feature particle- and light-like bands with parabolic and linear dispersions, respectively, similar to those found in monolayer transition metal dichalchogenide (TMD) semiconductors \cite{article:momentumArticle2}. Additionally, we find that excitons in BBG prefer to anti-align relative to the electric field. In turn, these rich excitonic properties, in conjunction with the voltage dependent band gap, yield highly \textit{in situ} customizable electronic and optical properties, with strongly temperature- and voltage-dependent emission properties. Our results are based on \textit{ab initio} modeling to describe BBG excitons with and without a finite center of mass momentum $\vec{Q}$ at varying electric fields between $30 \, \text{mV/Å}$ and $300 \, \text{mV/Å}$. As with recent experiments \cite{article:BBG4}, we limit our study to the case of vanishing net charge, i.e., equal displacement fields above and below the bilayer. A sketch of this setup can be seen in Fig. \ref{fig:introduction}a, with relevant angles and vectors shown on the right.

Recent work by Park \textit{et al.} on excitons in BBG show intriguing optical selection rules caused by pseudospin physics \cite{article:BBG1}. These authors utilize a BSE calculation, based on tight-binding band structures, with full rotational symmetry and no exchange coupling. Existing BSE calculations for BBG excitons are limited to the case of zero center of mass momentum $\Vec{Q}=0$ \cite{article:BBG1,article:BBG3}, justifying the omission of exchange coupling. However, exchange plays an important role for non-zero momentum BSE calculations due to the strong intervalley exchange coupling found in bright excitons at finite momentum \cite{article:LightlikeAndParticlelike}. 
Thus, to determine exciton band structures and momentum-resolved properties, we apply a highly accurate DFT-BSE approach outlined in App. \ref{sec:numDet}, including the exchange kernel and interlayer screening. The DFT band structures underlying the BSE are shown in Fig. \ref{fig:introduction}b, where the significant trigonal warping included in our model is clearly visible. The obtained momentum-resolved excitonic properties, namely the wave functions, energies, and optical matrix elements, will subsequently be used to describe light absorption and emission properties in the following sections. All presented results are for freely suspended BBG. However, screening due to sub- and superstrates can readily be incorporated, as shown in App. \ref{sec:numDet}, and, moreover, all conclusions are expected to remain qualitatively correct in the presence of such screening.

\section{Exciton susceptibility, ordering, and alignment}
A detailed picture of the optical properties in BBG requires accurate knowledge of the exciton band structures and optical transition strengths. As detailed in our previous work \cite{article:LightlikeAndParticlelike}, the rate of spontaneous emission is governed by two factors involving finite $\vec{Q}$: transition dipole moment and available phase space. In contrast, the $\vec{Q} = 0$ transition matrix elements $\vec{P}$, defined in Eq. \eqref{eq:momElems}, determine the sheet susceptibility to an excellent approximation, which in turn yields the optical absorption. Hence, before turning to exciton dispersions and emission in Sec. \ref{sec:momentum}, the remainder of this section shall present and analyze the results of $\vec{Q} = 0$ calculations, from which the peculiar exciton anti-alignment will also be discussed.

The DFT band structures, for states in the vicinity of the Dirac-point, are shown in Fig. \ref{fig:introduction}b. Here, the recognizable “Mexican hat”-valleys, with three distinct extrema in the $k$-space region arising from trigonal warping, are visible. These extrema are denoted mini-valleys, and the inter-mini-valley (IMV) vector between these $\vec{k}_{\text{IMV}}$. The mini-valleys play an important role for the brightness of the ground state exciton since its small, but non-zero, transition dipole moment is a consequence of the symmetry breaking induced by trigonal warping \cite{article:BBG1,article:BBG3,article:selectionRules}. In fact, without trigonal warping (i.e., with full rotational symmetry around $K$ and $K'$) the ground state exciton is completely dark \cite{article:BBG1, article:BBG3}. This is explained by the pseudospin characteristics of BBG, namely the two valleys will have an internal pseudospin angular momentum of $m_{\text{pseudo}} = -2 \, (+2)$ in the $K \, (K')$ valley \cite{article:ANDO1, article:BBG1}. In addition, exciton states possess an angular momentum due to their envelope function $m_{\text{env}}$. Thus, an analysis of the angular momenta using a continuum model with full rotational symmetry predicts the ground state exciton (1s, $m_{\text{env}} = 0$) to be dark, and the second exciton state (2p, $m_{\text{env}} = +1$ in $K$ and $m_{\text{env}} = -1$ in $K'$) to be bright for normal incident radiation, since the total angular momentum $m = m_{\text{pseudo}} + m_{\text{env}}$ must change by $\pm 1$ in optical transitions \cite{article:BBG1, article:BBG3}.

\begin{figure}[htp]
    \centering
    \includegraphics[trim={0cm 0cm 1.5cm 1cm},clip,width=\linewidth]{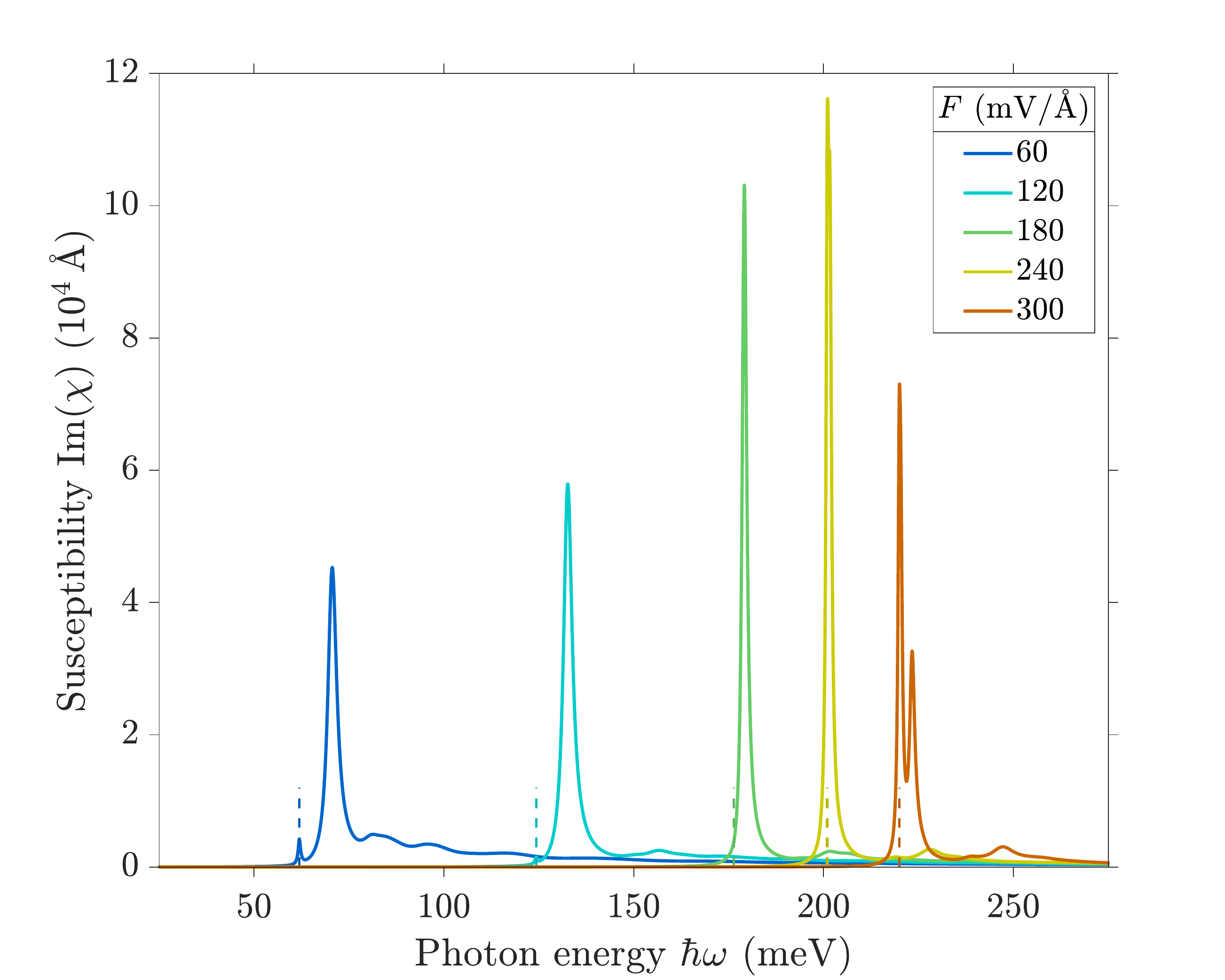}
    \caption{Imaginary parts of the in-plane sheet susceptibility $\text{Im}(\chi)$ vs. photon energy $\hbar \omega$ for BBG with external fields $F$ in the range $60-300 \text{mV/Å}$. The small dashed line segments indicate the energy of the ground state exciton.}
    \label{fig:Spektra}
\end{figure}

\begin{figure}[htp]
    \centering
    \includegraphics[trim={0cm 0cm 1.5cm 1cm},clip,width=\linewidth]{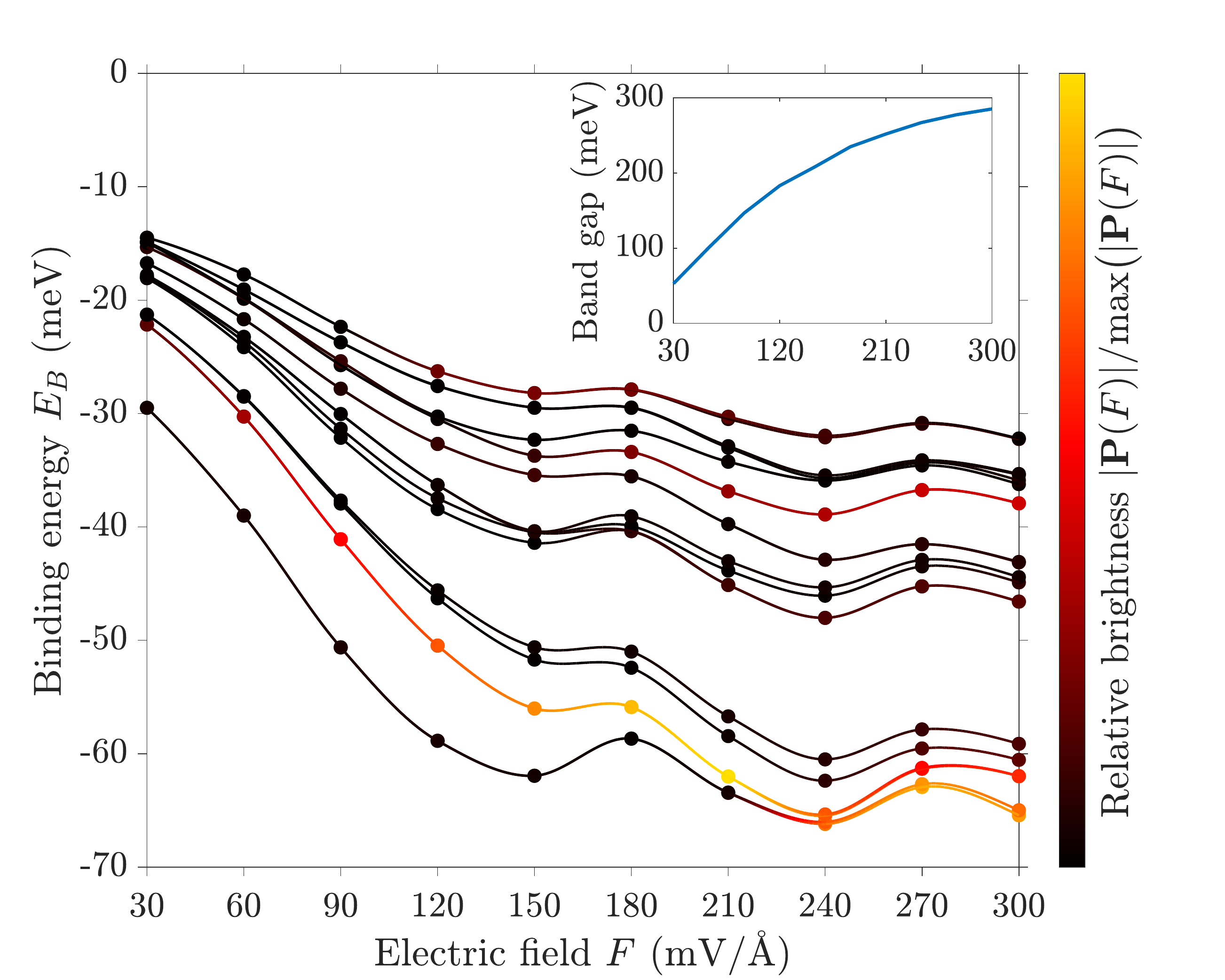}
    \caption{Exciton binding energies $E_B$ vs. electric field $F$ shown as dots with interpolating curves. The colors indicate the exciton brightness, determined by the transition matrix element $\vec{P}$, relative to the brightest overall state, among all the fields and bands. Inset: Scissor-shifted DFT band gap $E_g$ as a function of $F$.}
    \label{fig:XPlot}
\end{figure}

Optical absorption is described by the imaginary parts of the in-plane sheet susceptibilities shown in Fig. \ref{fig:Spektra} along with the ground state exciton energy. Here, the broadening parameter is taken to be very low, thus, the resonances reach very high values and the mostly dark ground state exciton can be observed similarly to experimental spectra \cite{article:BBG3}. Above the ground state exciton energy, the broadening parameter is set to increase linearly to match experimental findings \cite{article:BBG3}. For small external bias fields, the non-zero brightness of the mostly dark ground state is noticeable as a small peak at the ground state exciton energy. Furthermore, when the bias field increases, the energy difference between the first bright and the ground state exciton decreases, and above $240 \, \text{mV/Å}$ the states cross and the bright exciton becomes the ground state. This dependence is also visible in Fig. \ref{fig:XPlot} where the energy and brightness of the exciton states are shown as functions of the applied field. Interestingly, the binding energy does not seem to decrease monotonically with the field strength. This is caused by the mini-valleys becoming more prominent as the field increases. In turn, two field-dependent effects play a role for the binding energy: The mini-valleys localizing the exciton and the band gap dependent screening delocalizing it. Furthermore, the inset in Fig. \ref{fig:XPlot} shows that the calculated DFT band gap vs. applied field is a monotonically increasing function asymptotically approaching ${\sim}300 \, \text{meV}$, demonstrating the adjustable band gap of BBG.

\begin{figure}[htp]
    \centering
    \includegraphics[trim={0cm 0cm 1.25cm 1cm},clip,width=\linewidth]{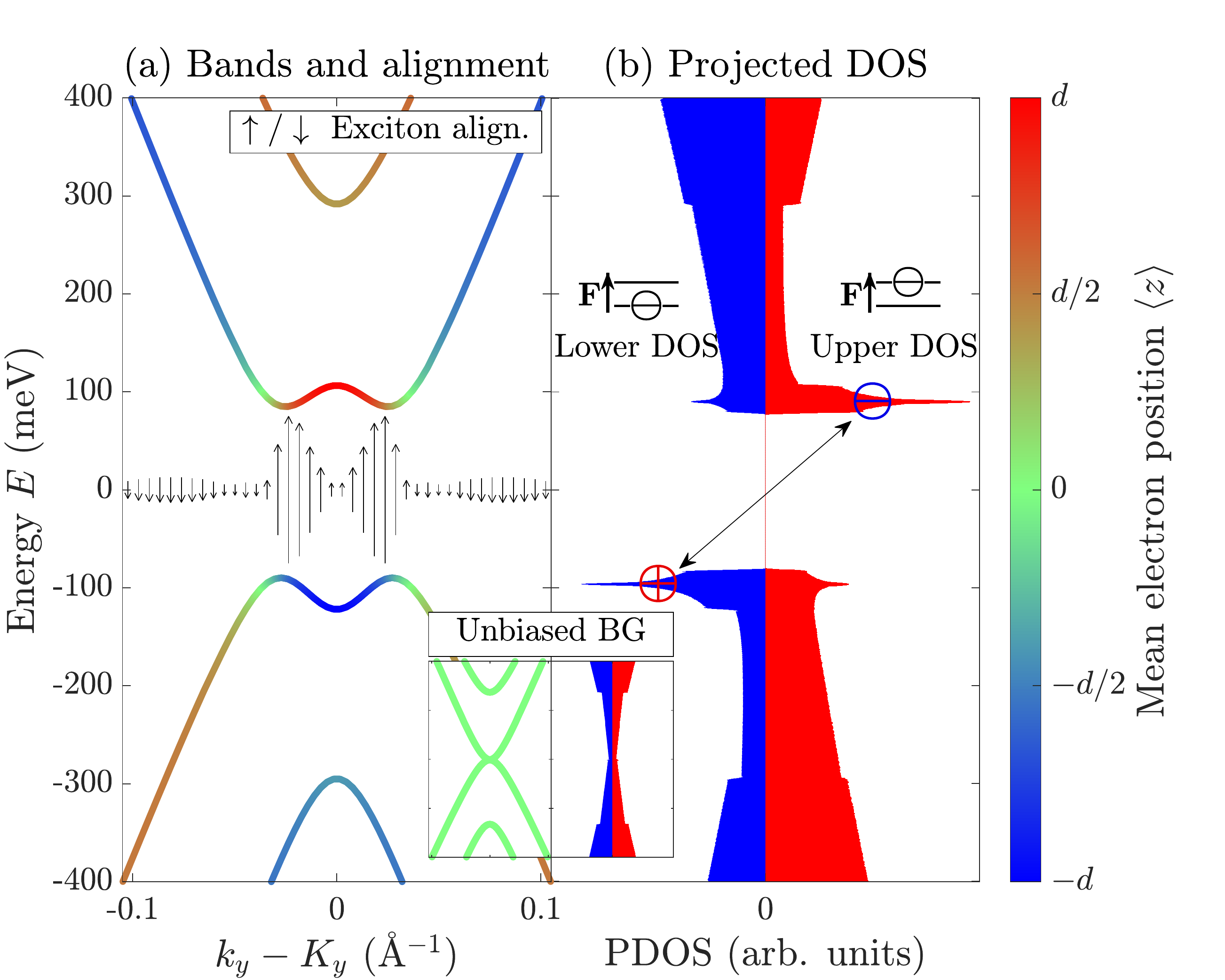}
    \includegraphics[trim={0.75cm 0cm 0.0cm 0.0cm},clip,width=\linewidth]{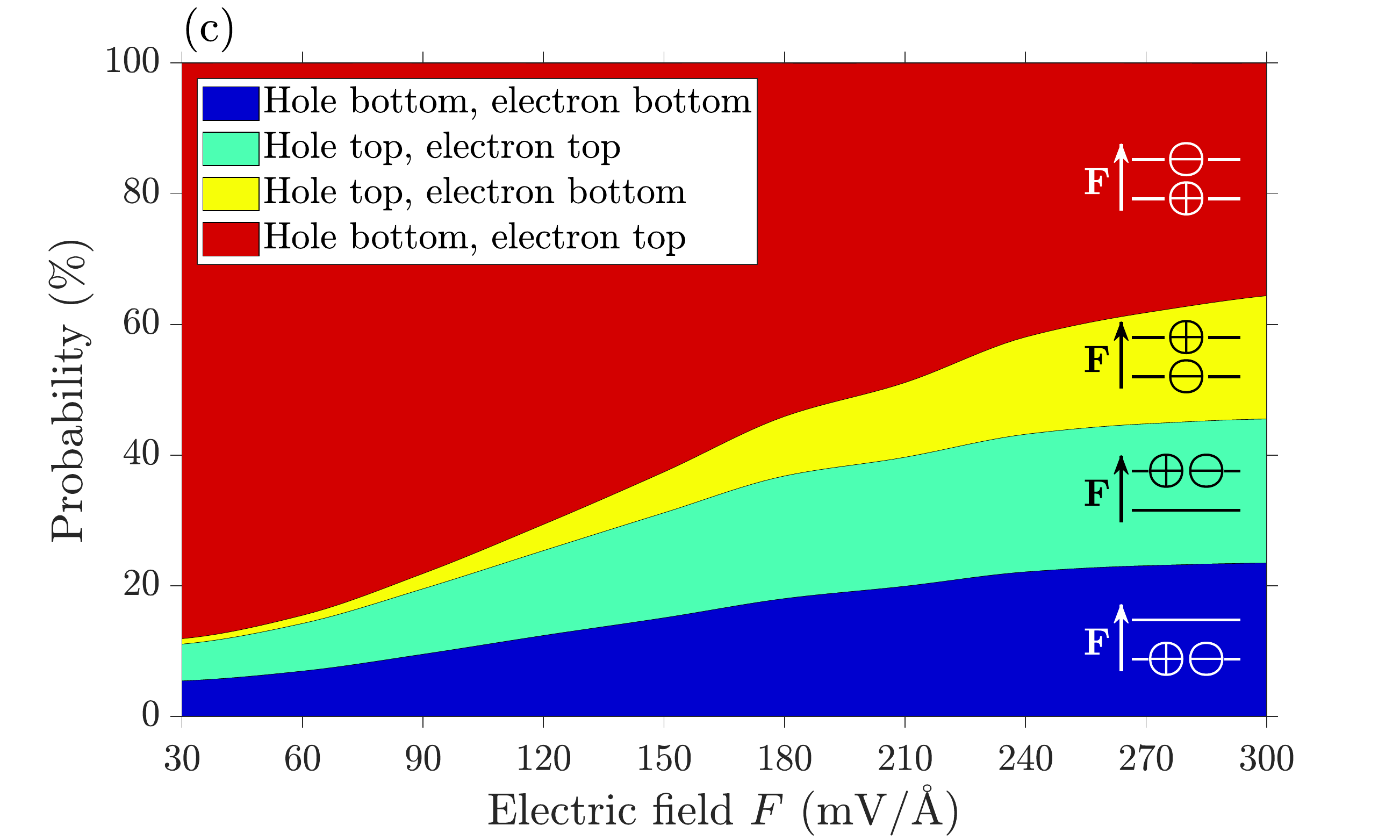}
    \caption{(a) Band structure and electron out-of-plane positional distribution for BBG with a field of $F = 180 \, \text{mV/Å}$. The line color represents the mean electron out-of-plane position $\left< z \right>$, with red (blue) being closer to the top (bottom) layer, located at $d = 1.74 \, \text{Å}$. The arrows in the band gap show the exciton alignment distribution $Z(\vec{k})$. (b) Density of states (DOS) for biased bilayer graphene projected onto the top or bottom layer. The inset in the bottom shows the band structure and electron projected DOS for unbiased bilayer graphene. (c) Probabilities of finding the four different exciton types with differing locations of electrons and holes vs. electric field $F$.}
    \label{fig:bandCross}
\end{figure}

Additionally, our calculations provide insight into the electron and hole positions for the exciton states, revealing quite unusual interlayer exciton dipole alignment due to the charge redistribution induced by the applied field.
In materials that are gapped in the absence of a bias, i.e. pristine semiconductors, excitons align with their dipole along the field \cite{article:Kamban_2020}.
In BBG, holes and electrons prefer to align with a dipole moment opposite to the applied external field corresponding to the anti-aligned exciton shown in Fig. \ref{fig:introduction}a. This anti-alignment has been observed in other works \cite{article:BBG1}, however, not analyzed to the extent presented here. Naturally, as the applied field increases, this effect weakens as the energy cost increases due to the energetically unfavorable dipole alignment.

This peculiar exciton dipole alignment may seem counter-intuitive at first, however, it is qualitatively a consequence of the electron distribution due to the bias in the out-of-plane direction. 
In more detail, the physics behind anti-alignment may be understood from Fig. \ref{fig:bandCross}, which shows the band structure, electron distribution, projected density of states, and partial exciton alignment of BBG.
In particular, in the ground state, i.e. a filled valence band and no excitons, the electrons position themselves such that the net internal electric field counteracts the external one, similarly to what happens in a conductor \cite{article:BBG8}. 
This is evident from the asymmetric distribution of the states near the band gap. Consequently, the lowest energy direct transitions will leave a hole in the bottom layer and add an electron in the top layer for an external field in the positive out-of-plane direction.
This is particularly noticeable in the projected density of states in Fig. \ref{fig:bandCross}b, where a van Hove singularity is observed near the band gap due to the saddle points arising from the "Mexican hat" valleys.
Accordingly, a significant density of anti-aligned transitions is found near the band gap, such that the exciton wave function primarily consists of these, with a few normal-aligned transitions on the sides of the “Mexican hat” valley in Fig \ref{fig:introduction}b.
This is illustrated by the exciton alignment distribution, shown by the arrows in Fig \ref{fig:bandCross}a.
We define this distribution as $Z(\vec{k}) = (\left< z_c \right> - \left< z_v \right>) \abs{A_{\vec{k} c v}^\lambda} \norm{\vec{k} - \vec{K}}$, where $c$ and $v$ are chosen as the lowest conduction and highest valence band, respectively, and $A_{\vec{k} c v}^\lambda$ is the wave function of exciton $\lambda$ in $k$-space, see Eq. \ref{eq:excitonWFS}. Here, the $\norm{\vec{k} - \vec{K}}$ weight arise from the circumference of the circle centered at $K$ (applying the approximate circular symmetry around $K$).
As such, $Z(\vec{k})$ shows the relative contribution from different $k$-points to the total exciton alignment.

Quantitatively, the probabilities for the different exciton alignments are shown in Fig. \ref{fig:bandCross}c for different fields. Here, it can be seen that, especially for the lower applied fields, the excitonic state is almost entirely anti-aligned. The probabilities shown are all for the ground state exciton, however, similar probabilities are observed for all higher energy bound states. The probabilities are calculated by integrating the electron probability densities shown in Fig. \ref{fig:wfs}, as detailed in App. A.3.


In high and intermediate fields, the excitons will have large electron-hole overlaps compared to the near pure interlayer excitons found at very low fields. This is because the exciton states at high and intermediate fields are mixed superpositions of aligned and anti-aligned transitions, as evident from Fig. \ref{fig:bandCross}c and \ref{fig:wfs}, and the probability of finding the electron and hole in the same layer is significantly increased. This also, partly, explains why the excitons at higher fields have larger transition matrix elements, as seen in Fig. \ref{fig:XPlot}, as this requires large electron-hole overlaps.





\begin{figure*}[htp]
    \centering
    \includegraphics[trim={7cm 1cm 5cm 2cm},clip,width=\linewidth]{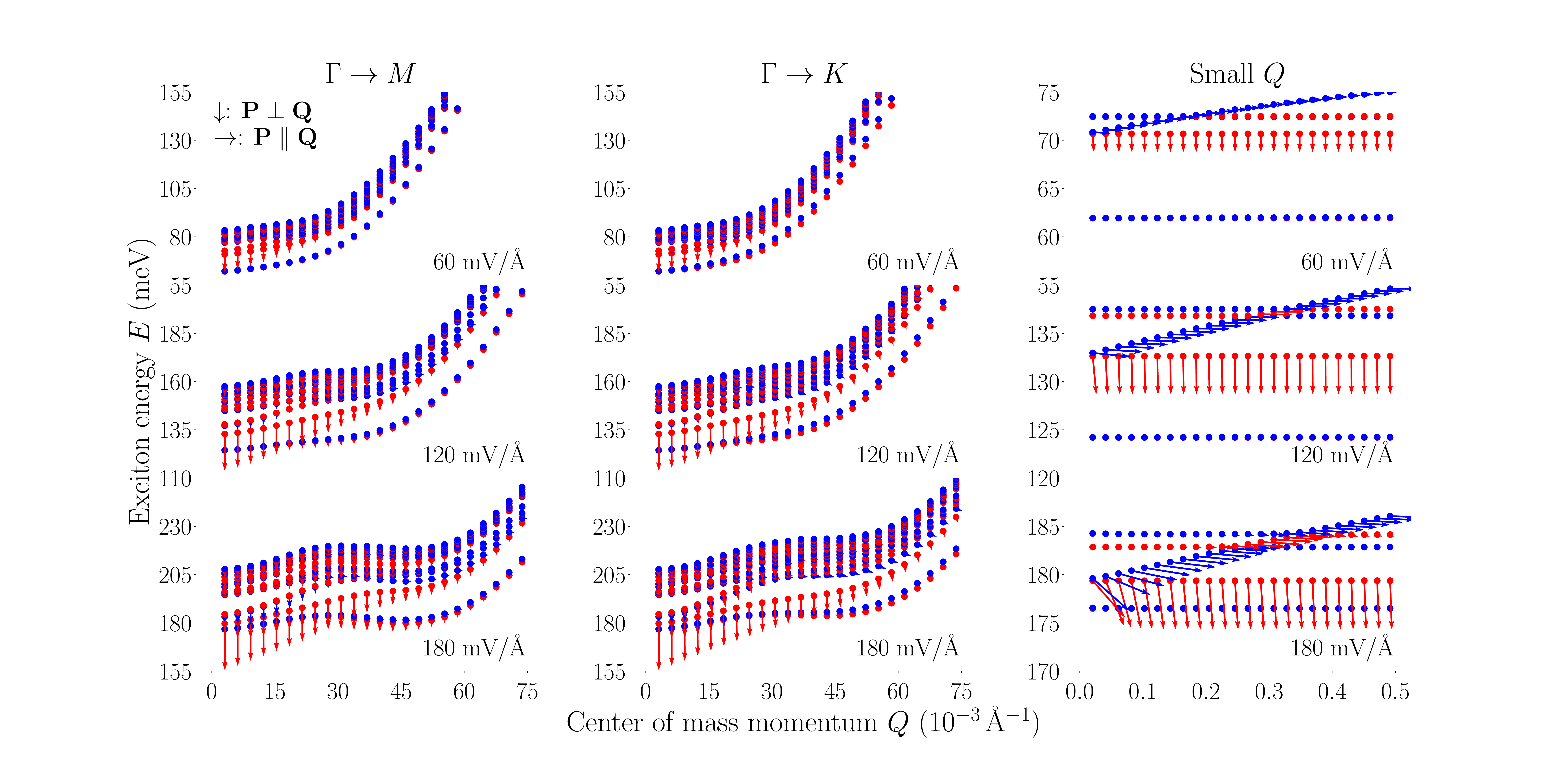}
    \caption{Exciton band structure for the twenty lowest excitons in BBG, with the magnitude of the exciton transition matrix element projected upon $\vec{Q}$, shown as a vector with components corresponding to either parallel or perpendicular projection. Rows correspond to different external fields and columns to different $\vec{Q}$ paths. In the third column, $\vec{Q}$ is along the $\Gamma \rightarrow M$ line, however, any other direction yields similar bands, as BBG is practically isotropic for small $\vec{Q}$.}
    \label{fig:energies}
\end{figure*}

\begin{figure}
    \centering
    \includegraphics[trim={0cm 0cm 0cm 1cm},clip,width=\linewidth]{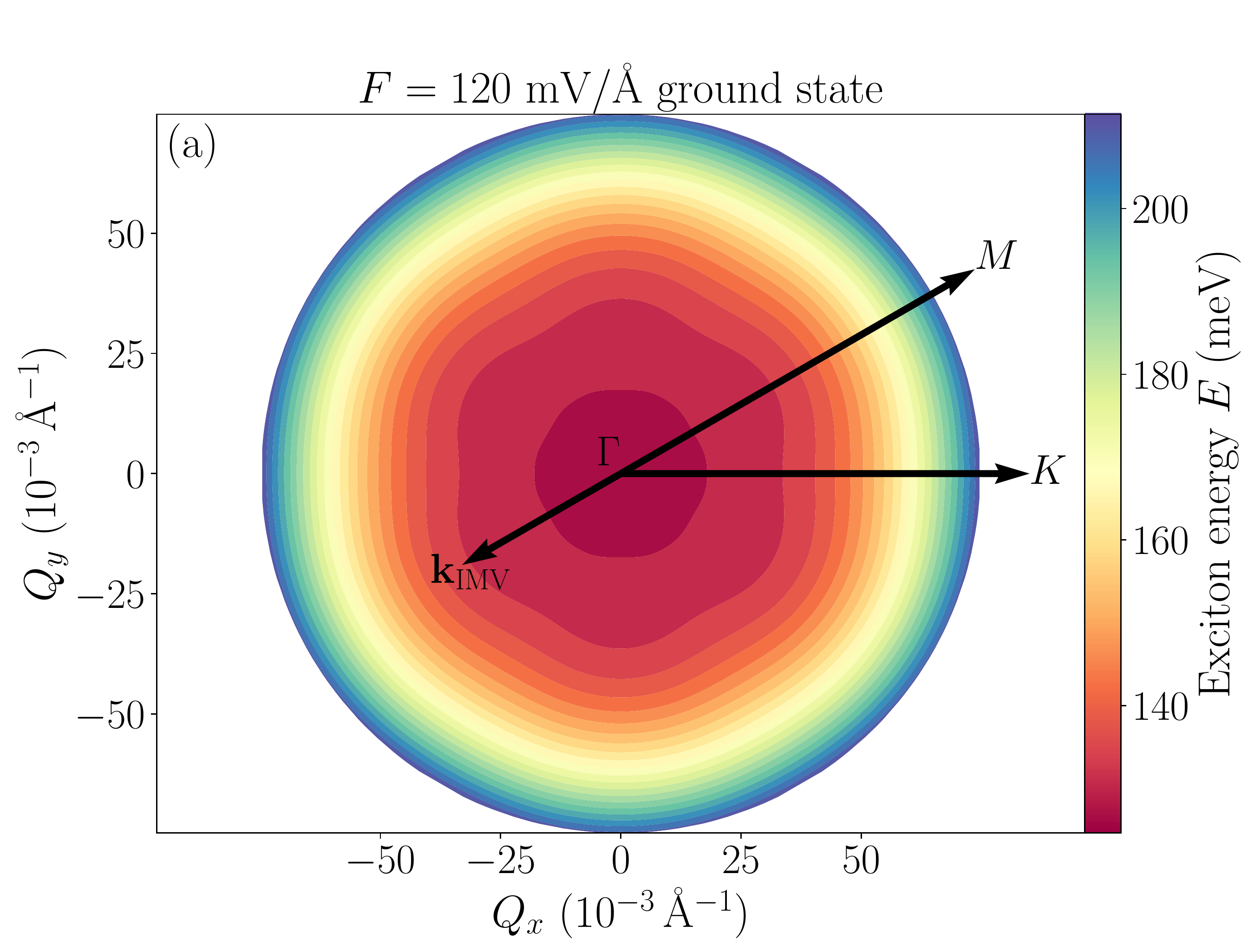}
    \includegraphics[trim={0cm 0cm 0cm 1cm},clip,width=\linewidth]{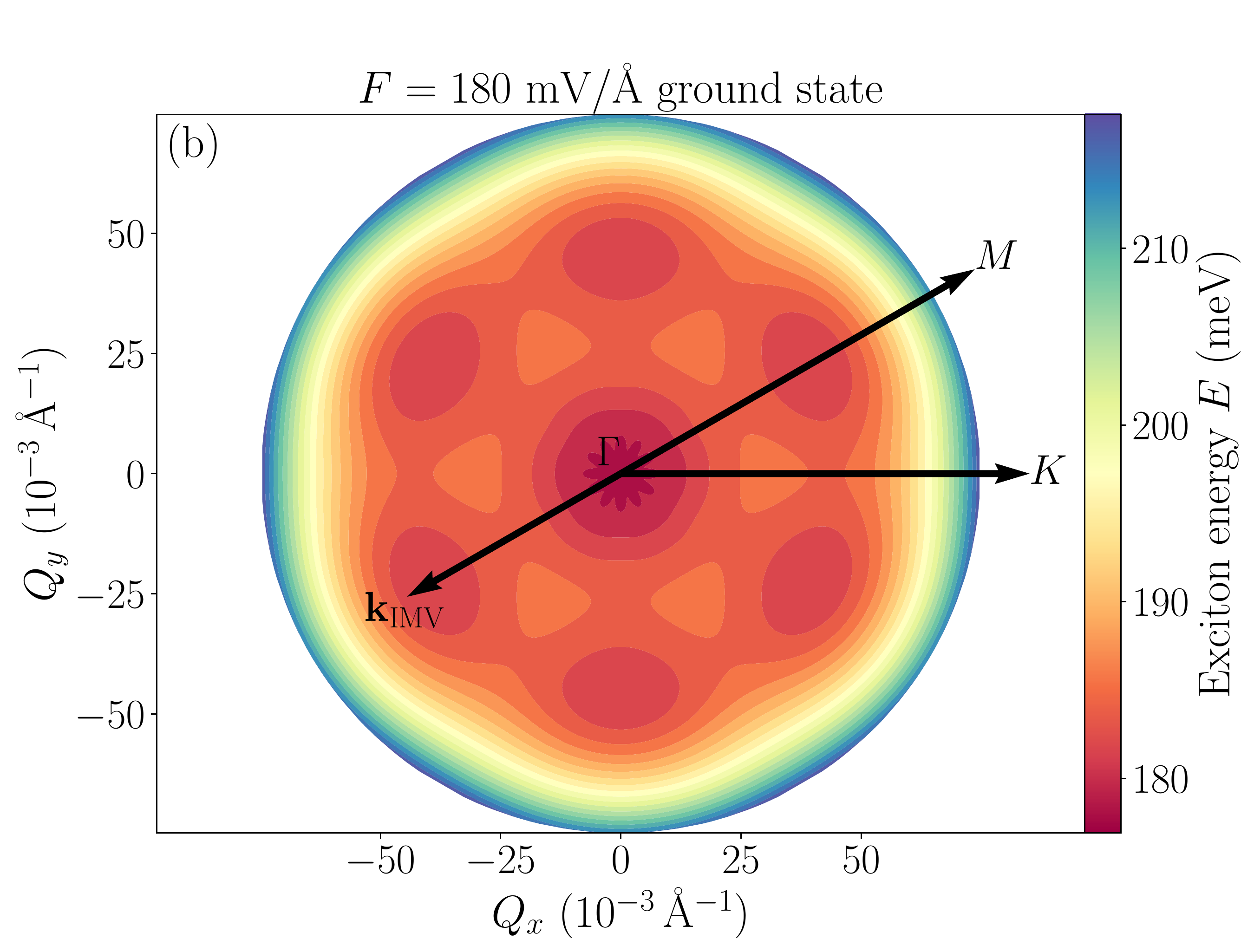}
    \caption{Surface plot of the exciton bands for the ground state exciton of BBG, in the $(Q_x,Q_y)$ plane, at (a) $F = 120 \, \text{mV/Å}$ and (b) $F = 180 \, \text{mV/Å}$.}
    \label{fig:2dfig}
\end{figure}

\section{Exciton dispersion and optical emission}
\label{sec:momentum}
We now turn to exciton band structures and emission properties, requiring finite $\vec{Q}$. In particular, the rate of spontaneous emission depends on the momentum resolved exciton transition matrix element $\vec{P}$ and available phase space. 
Due to energy and momentum conservation, the photon frequency $\omega$ is given as $\hbar \omega = E(\vec{Q})$, with $E(\vec{Q})$ the exciton energy, and the in-plane photon momentum is given as $\vec{q}_{x,y} = \vec{Q}$. Consequently, the available phase space for emission is limited by the photon momentum $\omega/c$, and only excitons with $\abs{\vec{Q}} \le \omega / c$ may decay radiatively.
Additionally, phase space becomes even more restricted when considering photon transversality. This requirement means that emission rates are proportional to $\sin^2\varphi$, where $\varphi$ is the angle between $\vec{q}$ and $\vec{P}$ \cite{article:LightlikeAndParticlelike}, see Fig. \ref{fig:introduction}a. Hence, exciton dispersions are required in order to obtain the spontaneous emission properties.

To obtain exciton dispersions for BBG, we include a center of mass momentum $\vec{Q}$ in the BSE. The resulting exciton bands are shown in Fig. \ref{fig:energies}, where three different $\vec{Q}$ paths are presented for different field strengths. On the large $\vec{Q}$-scale applied in the first and second columns, exciton bands are primarily characterized by the mini-valleys inherited from the DFT band structure. Additionally, except for the bright exciton, they show two-fold degeneracy for $\vec{Q}$ in the vicinity of $\Gamma$. This is in agreement with the seemingly flat bands seen in the third column, where the momentum scale is significantly lower than the inter-mini-valley distance. At $\vec{Q}=0$, the lowest (doubly degenerate) exciton is dark, while the second-lowest exciton is bright. Hence, the third column shows that the bright exciton splits into a light-like (linear) and particle-like (parabolic) band for finite $\vec{Q}$. Note that the curvature of the parabolic bands is small, thus, these bands are seemingly flat. Furthermore, the components of $\vec{P}$ parallel and perpendicular to $\vec{Q}$ are shown by arrows in the plot. Hence, the transition matrix elements of the light-like band are approximately parallel to the $\vec{Q}$, while those of the particle-like band are perpendicular to $\vec{Q}$. This is similar to the case of monolayer TMDs and is a consequence of the hexagonal symmetry shown by BBG and TMDs \cite{article:LightlikeAndParticlelike,article:momentumArticle2}. 

The reason only bright excitons split into light- and particle-like bands derives from the momentum-dependence of the intervalley exchange. 
Previously, we established an effective model describing the band dispersion, including inter- and intravalley exchange, and spin-orbit coupling \cite{article:LightlikeAndParticlelike}. However, in BBG, the spin-orbit coupling is negligible. Furthermore, only the intervalley exchange has terms linear in $Q$ \cite{article:LightlikeAndParticlelike, article:momentumArticle2}. Consequently, a model including first-order terms in $Q$ predicts a splitting of $2|\braket{K|v_x|K'}|$ between the light- and particle-like bands, with intervalley exchange 
\begin{equation}
    \braket{K|v_x|K'} \propto \frac{1}{Q}  (\vec{Q} \cdot \vec{p}_{\vec{K}}^*) (\vec{Q} \cdot \vec{p}_{\vec{K'}}),
    \label{eq:exc_propto}
\end{equation}
where $\vec{p}_{\vec{k}}=\braket{v_{\Vec{k}}|\vop{p}(0)|c_{\Vec{k}}}$ is the interband transition matrix element in the long-wavelength approximation \cite{article:LightlikeAndParticlelike}, with $\vop{p}(0)$ defined in Eq. \eqref{eq:momOp}. This means that large transition matrix elements are required for noticeable light-like bands to appear. Consequently, the lowest doubly degenerate dark band, with a nearly vanishing transition matrix element, does not split. In contrast, the light-like band of the bright exciton has a substantial slope. Additionally, the approximately parallel or perpendicular projection of the transition matrix element to $\vec{Q}$, for the light- and particle-like bands, can also be naturally explained within this model \cite{article:LightlikeAndParticlelike}.

Next, we turn to the field dependence of the exciton dispersions. As evident from the upper rows in Fig. \ref{fig:energies}, the dependence of the energy $E$ on the angle of $\vec{Q}$ is very weak for small applied electric fields. However, clear fingerprints of mini-valleys appear when applying greater fields. This can be seen from the local minima observed in the first and second columns for $180 \, \text{mV/Å}$ around $Q = 50 \times 10^{-3} \, \text{Å}^{-1}$, where the valley is noticeably deeper along the $\Gamma \rightarrow M$ direction. This is particularly evident in Fig. \ref{fig:2dfig}, where the angle-resolved dispersion is plotted. Furthermore, the inter-mini-valley distance $\vec{k}_{\text{IMV}}$ is shown and matches perfectly with valleys in the exciton dispersion, showing clearly the effect of the significant trigonal warping. However, it is clear that for $\norm{\vec{Q}} \ll \norm{\vec{k}_{\text{IMV}}}$, the angular dependence remains negligible. Consequently, the spontaneous emission will show no dependence on the azimuthal angle and the calculation of spontaneous emission can be reduced in dimensionality \cite{article:LightlikeAndParticlelike}.


\begin{figure}[htp]
    \centering
    \includegraphics[trim={1cm 0cm 1cm 1cm},clip,width=\linewidth]{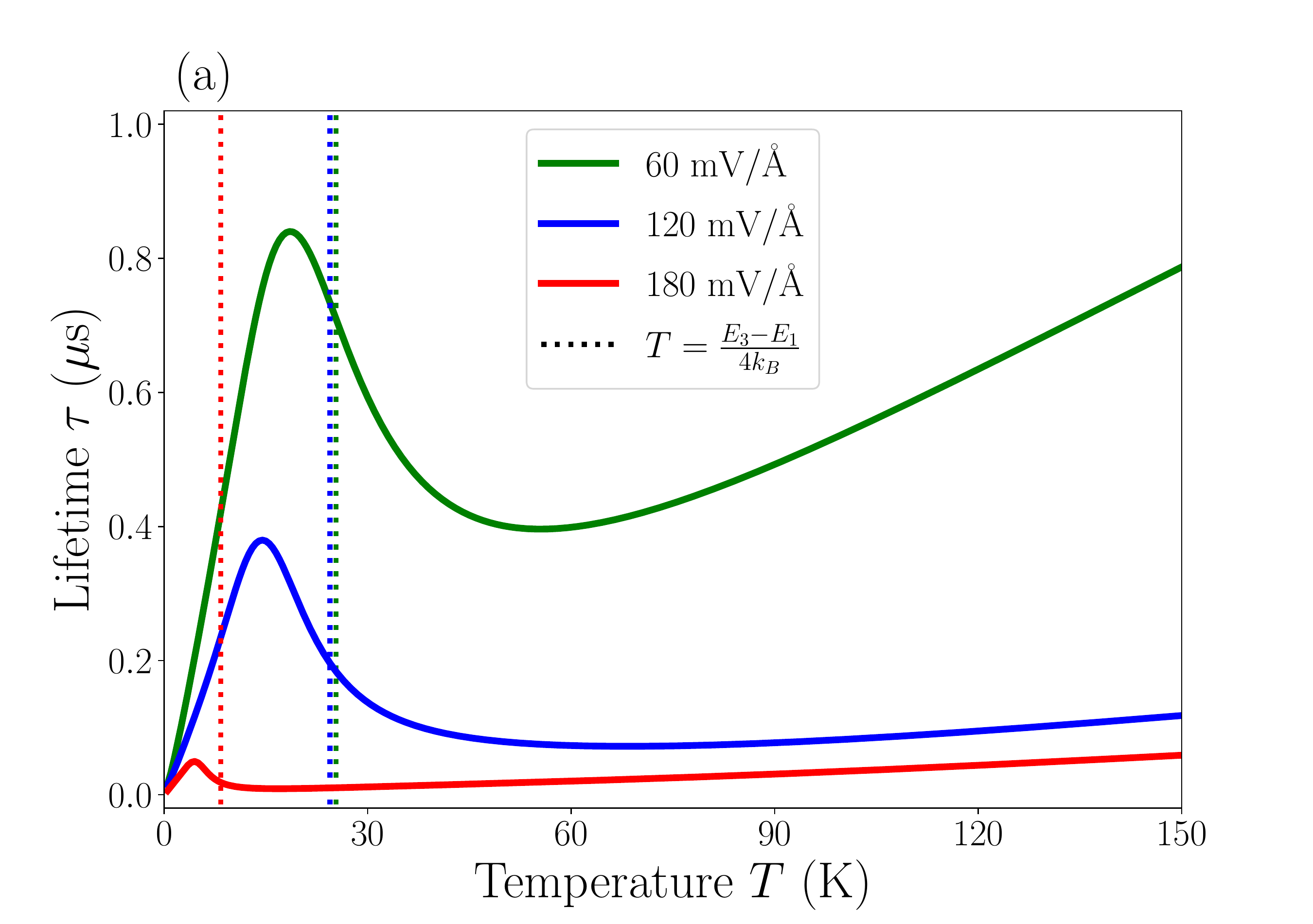}
    \includegraphics[trim={1cm 3cm 1cm 4cm},clip,width=\linewidth]{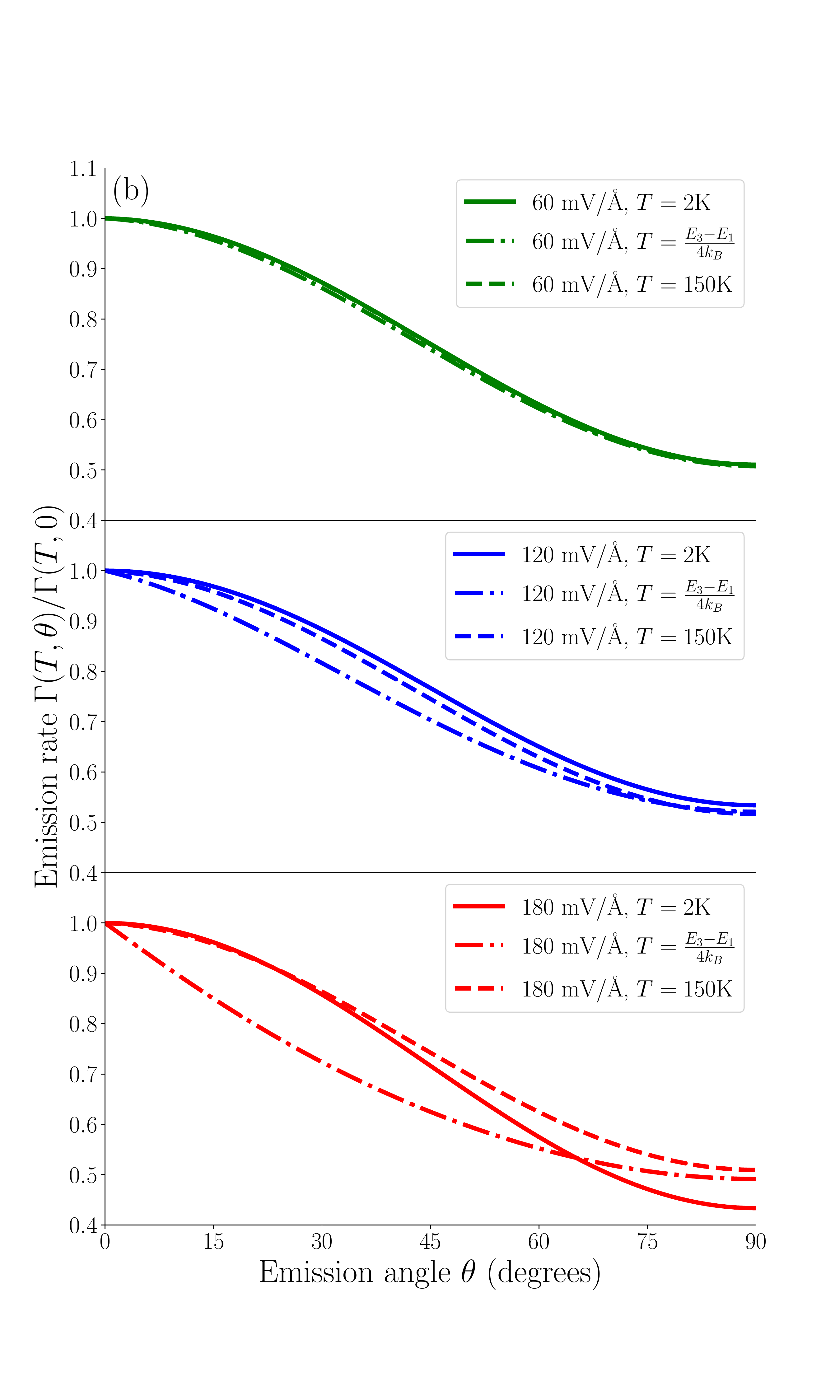}
    \caption{(a) Radiative lifetime $\tau$ vs. temperature $T$ at fields $60$, $120$, and $180 \, \text{mV/Å}$. Dotted lines indicate the temperature onset for significant population of the bright exciton bands, at which the lifetime drops. Here, $E_1$ and $E_3$ are energies of the ground state and first bright exciton (first and third eigenvalue), respectively. (b) Normalized emission  $\Gamma(T, \theta)/\Gamma(T, 0)$ as a function of the  inclination angle $\theta$, at $T = 2 \, \text{K}$, $T = 150 \, \text{K}$, and the temperature onset for population of the bright bands.}
    \label{fig:emission}
\end{figure}

Radiative lifetimes are modelled using the approach applied previously \cite{article:LightlikeAndParticlelike}. Briefly, we utilize a momentum-resolved expression for the radiative lifetimes of 2D materials, derived from Fermi's golden rule \cite{article:lifetimeEquation}. Subsequently, we thermally average this expression using a Boltzmann distribution to obtain the temperature dependent radiative lifetime. To reduce numerical error in the low-temperature region, we approximate the transition matrix elements as constant $\vec{P}(\vec{Q}) \simeq \vec{P}(\frac{\omega}{2c} \unvec{Q})$, and the particle- and light-like bands as flat and linearly increasing, respectively, for $Q \le \omega/c$. The approximations improve numerical stability \cite{article:LightlikeAndParticlelike}. In Fig. \ref{fig:emission}, the calculated radiative lifetimes as functions of temperature can be seen, as well as angular emission profiles. The radiative lifetimes are strongly dependent on temperature and applied electric field, with lifetimes in both the nano- and microsecond range. A significant factor in this dependence is the thermal occupation of the bright exciton state. In particular, this is due to the energy difference between the ground state excitons (1 and 2) and first bright excitons (3 and 4). Consequently, at low thermal energies, primarily the almost dark ground state exciton band is occupied. However, at higher temperatures, the probability of occupying bright higher energy states increases. This is evident from the local maximum in the radiative lifetime $\tau(T)$ curves, right before the temperature onset for significant bright exciton population. This onset is $T = \frac{E_3 - E_1}{4 k_B}$, where $E_3$ is the energy of the first bright exciton and $E_1$ the energy of the ground state exciton, as indicated in Fig. \ref{fig:emission}a. Upon increasing the temperature even further, a local minimum is observed. This minimum corresponds to the temperature, where the maximum population of bright states in the available phase space occurs. At even higher temperatures, the lifetimes increase linearly, a consequence of the thermal energy increasing the population of states outside the available phase space. The angular spectra in Fig. \ref{fig:emission}b also reveal a temperature dependence, namely, as the first bright excitons become significantly populated, the distinct angular profile of a partly populated light-like exciton becomes visible. That is, for the light-like exciton, due to its linear dispersion, only $\vec{Q} \approx 0$ states are populated. Consequently, because of in-plane momentum conservation, the relative probability of emission with photon momentum $q_z \approx \omega/c$ (light perpendicular to the plane) increases. This is particularly noticeable in the $F = 180 \, \text{mV/Å}$ case due to the very low energy difference between the ground state and the bright exciton.

\section{Summary}
In summary, we have studied the absorption, band structure, and optical emission properties of excitons in BBG subjected to out-of-plane electric fields in the range $F=30-300 \, \text{mV/Å}$, using a first-principles DFT+BSE approach. 
Our results provide a detailed description of the field-dependent ordering of exciton states and absorption properties, the anti-aligned nature of the excitons, the rich exciton band structure, and the strongly field- and temperature-dependent emission properties. 

We find the ground state exciton to be dark for low fields, however, the ordering of excitons changes in high fields such that at $F \approx 240 \, \text{mV/Å}$ the lowest bright exciton becomes the ground state exciton.
Furthermore, excitons in low fields are almost exclusively anti-aligned with respect to the applied field. This is due to bias-induced charge transfer between the layers, leading to a dipole aligned along the field in the ground state. This dipole is induced by electron states near the valence band top, thus, excitons formed by holes near the valence band top and electrons near the conduction band bottom tend to have a dipole aligned opposite that of the ground state. At higher fields, however, the probability of excitons being anti-aligned drops, due to the energetically unfavorable dipole alignment, but remains the dominant alignment even at $F = 300 \, \text{mV/Å}$.

When considering excitons with a finite center of mass momentum $\vec{Q}$, we find that, due to the hexagonal symmetries of BBG, the bright exciton state splits into light- and particle-like bands with transition matrix elements parallel and perpendicular to $\vec{Q}$, respectively. Importantly, a requirement for this split is the large transition matrix element observed for the bright exciton.
Finally, based on the calculated exciton bands, we obtain radiative lifetimes as a function of temperature. In particular, we find that the energy difference between the ground state exciton and first bright exciton plays a major role in the radiative lifetimes and their temperature dependence. Hence, by changing the bias field, this energy difference can be modified, yielding highly customizable emission properties for BBG. 


\begin{acknowledgments}
The authors are supported by the CNG center under the Danish National Research Foundation, project DNRF103. 
\end{acknowledgments}

\appendix
\renewcommand{\thesubsection}{\Alph{section}.\arabic{subsection}}
\section{Numerical details}
\label{sec:numDet}
\begin{figure}[htp]
    \centering
    \includegraphics[trim={1cm 1.2cm 2cm 2cm},clip,width=\linewidth]{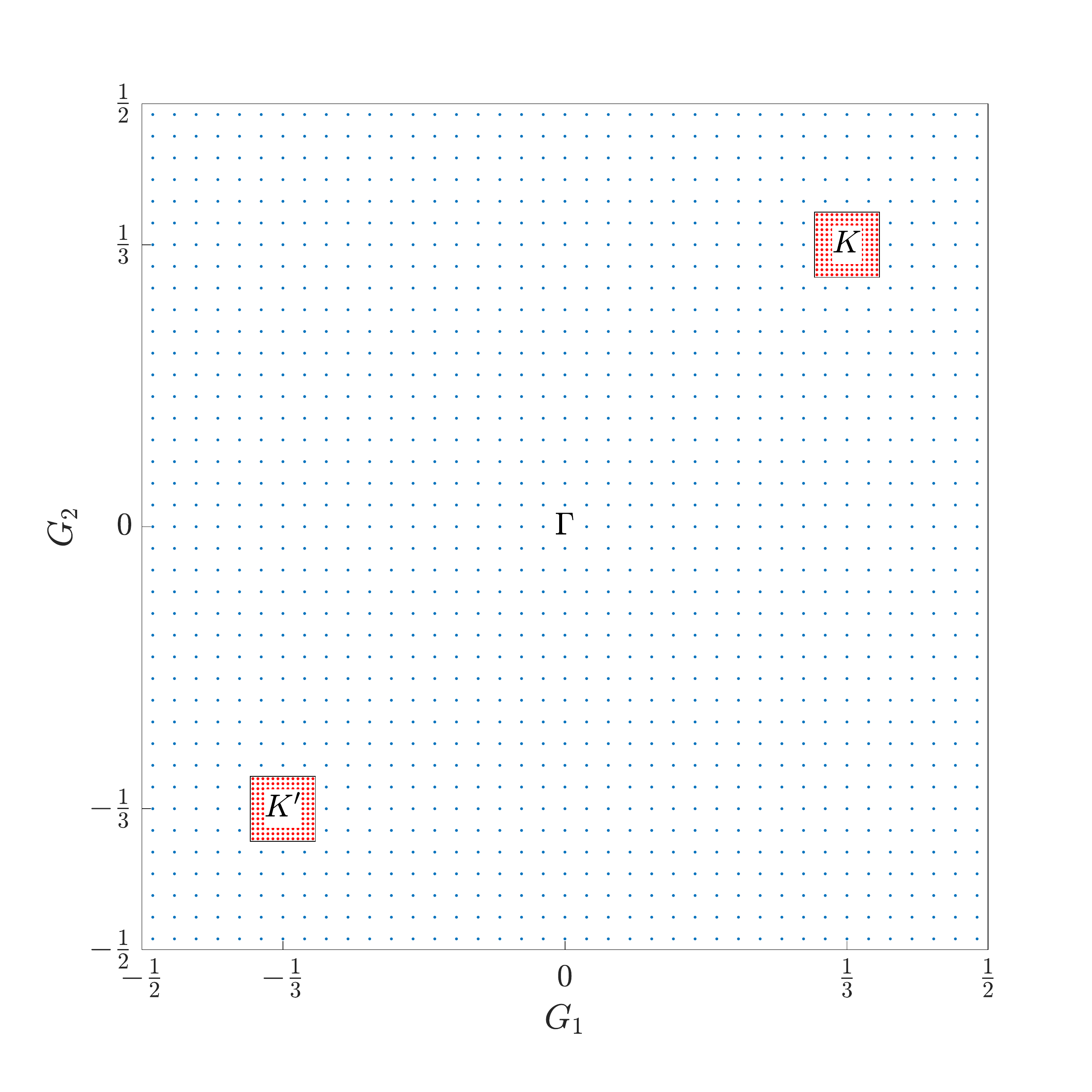}
    \caption{Sketch of the applied $k$-grid in the BSE and self-consistent DFT calculation with low-density (blue-dotted) and high density (red-dotted) sections. The coordinates are given in reciprocal basis vectors $\Vec{k} = G_1 \Vec{G}_1 + G_2 \Vec{G}_2$.}
    \label{fig:BZ}
\end{figure}

First-principle BSE exciton calculations for BBG are numerically demanding due to the very long screening lengths and non-parabolic dispersions around the $K$ and $K'$ symmetry points. Furthermore, attaining a $k$-space resolution corresponding to optical momenta is required for calculating radiative lifetimes. For this reason we apply a non-uniform $k$-grid as illustrated in Fig. \ref{fig:BZ}, which is very dense around the $K$ and $K'$ symmetry points, in all steps of our calculation. Additionally, we apply a computationally efficient approach, in which a limited set of bands is retained and interpolation between $\Vec{Q}$-points is applied to acquire exciton energies and transition matrix elements. Briefly, our calculation consists of three main steps, namely, the self-consistent DFT calculation, the interpolant-grid DFT calculation, and the BSE calculation. Here, the DFT calculations are carried out using GPAW \cite{article:GPAW1, article:GPAW2, article:GPAW3, article:GPAW4, article:GPAW5} with PBE exchange and correlation.

The bilayer graphene geometry is found by relaxing the structure in the absence of an external field, i.e., $F=0$, from which an interlayer distance of $2d = 3.48 \text{Å}$ and a lattice constant of $a = 2.46 \text{Å}$ are found, corresponding well with experimental findings \cite{article:bilayerDistance}. Furthermore, to eliminate unphysical coupling between out-of-plane repeated layers, a vacuum slab of $20 \text{Å}$ is applied. First, in the self-consistent DFT calculation, we apply a $k$-grid density of ${\sim} 746 \, \text{Å}^{2}$ in the low-density region and ${\sim}1.68 \times 10^5 \, \text{Å}^{2}$ in the high-density regions encompassing a $1.2 \times 10^{-2} \, \text{Å}^{-2}$ area around the $K$ and $K'$ symmetry points. Second, we determine the screening length and dense interpolant-grid based on the self-consistent calculation. Subsequently, the energies are scissor-shifted to a fit of existing experimental band gap data \cite{article:BBG4}. The interpolant grid consists of ${\sim}3.3 \times 10^4$ $k$-points, distributed non-uniformly with a density following a continuous function, namely a double generalized normal distribution, with higher densities closer to $K$ and $K'$, ensuring accurate interpolation. Finally, the BSE is solved on a grid with a $k$-point density of ${\sim} 119 \, \text{Å}^{2}$ in the low density region and ${\sim}5.3 \, \times 10^4 \text{Å}^{2}$ in the high density regions.

The BSE is constructed from two valence and two conduction bands, disregarding spin-orbit coupling such that only one bright singlet state is found for every spinor. This has been shown to be a good approximation when the intravalley exchange is significantly larger than the SO-coupling \cite{article:LightlikeAndParticlelike}. We apply a bilayer screening approach for the Coulomb interaction, in which the interlayer separation is included as shown in App. A.1 and A.2, similarly to the approach by Kamban \textit{et al.} \cite{article:Kamban_2020}. Furthermore, only external screening is included for the exchange term, thus, it is given as $w_x(\Vec{q}) = 2 v_{\text{bare}}(\Vec{q})/(\eps_a + \eps_b)$, where $\eps_a$ and $\eps_b$ are the dielectric constants of the super- and substrate, respectively \cite{article:exchangeScreening,article:Louie_2021_exchange}. The center of mass momentum $\Vec{Q}$ is handled by adding half to the conduction state and subtracting half from the valence state, such that $c_{\Vec{k}}$ is replaced by $c_{\Vec{k}+\Vec{Q}/2}$ and the valence states $v_{\Vec{k}}$ by $v_{\Vec{k}-\Vec{Q}/2}$. This ensures that the $K$-point stays as centered as possible in the dense part of the $k$-grid. Due to the non-uniform $k$-grid either a new DFT calculation must be made or an interpolation scheme applied for every $\vec{Q}$ as chosen in the present work. In particular, we obtain the $k$-grid shifted by $\Vec{Q}/2$ using a 3rd-degree polynomial spline scheme, ensuring any momentum $\Vec{Q}$ can be resolved. The transition matrix element of exciton $\lambda$ is calculated as
\begin{equation}
    \vec{P}^{\lambda} = \frac{\sqrt{2} A}{4 \pi^2}\sum_{v,c} \int_{\text{BZ}} A_{\vec{k}cv}^{\lambda}(\vec{Q}) \braket{v_{\Vec{k} - \Vec{Q}/2}|\vop{p}(\vec{q})|c_{\Vec{k} + \Vec{Q}/2}} d^2 \vec{k},
    \label{eq:momElems}
\end{equation}
where $A_{\vec{k}cv}^{\lambda}(\vec{Q})$ is the exciton eigenvector element for the corresponding valence-conduction band combination, $\Vec{q}$ is the photon momentum, the factor of $\sqrt{2}$ arises due to spin, and $\vop{p}(\vec{q})$ is the momentum operator, given as
\begin{equation}
    \vop{p}(\vec{q}) = -\frac{i \hbar }{2} \left\{ \nabla \exp{-i \Vec{q} \cdot \Vec{r}}  + \exp{-i \Vec{q} \cdot \Vec{r}} \nabla \right\}.
    \label{eq:momOp}
\end{equation}
It should be noted that the BSE calculations for BBG with electric fields above $180 \, \text{mV/Å}$, as shown in Fig. \ref{fig:Spektra} and \ref{fig:XPlot}, use a slightly lower $k$-point density, however, due to the larger inter-mini-valley distance the impact on convergence is minimal. 
\onecolumngrid
\subsection{Bilayer Coulomb interaction}
\label{sec:screening}
In bilayer structures, both intra- and interlayer Coulomb interaction are present. To describe the screening, we shall establish a simple model accounting for the distance between the hole and electron in both cases, assuming the two layers to be identical. We approximate the dielectric function as piecewise constant
\begin{equation}
    \varepsilon(q, z) = 
    \begin{cases}
    \eps_a, & z>2d \\
    \eps, & -2d \le z \le 2d \\
    \eps_b, & z<-2d
    \end{cases},
\end{equation}
where $2d$ is the effective thickness of one graphene layer and thus also the distance between the two layers, $\eps$ the dielectric function of BBG, and $\eps_a$ and $\eps_b$ the dielectric functions of the super- and substrate, respectively. Following Kamban \textit{et al.} \cite{article:Kamban_2020}, however, using that both layers are of identical material, dielectric function $\eps$, and width $2d$, we find for the Coulomb potential
\begin{equation}
    w(q, z, z') = \frac{e^2}{2 \eps_0 q} 
    \begin{cases}
    C \exp{-q z} & z > 2d \\
    \eps^{-1} \exp{-q \abs{z-z'}} + A \exp{- q z} + B \exp{q z} & -2d \le z \le 2d \\
    D \exp{q z} & z < -2d
    \end{cases}.
\end{equation}
Applying the appropriate boundary conditions and solving the corresponding system of equations with both the hole and electron coordinate restricted to $\abs{z}, \abs{z'} < 2d$, we get
\begin{equation}
    w(q, z, z') = \frac{e^2 \exp{-q \abs{z + z' + 4d}}}{2 q \eps \eps_0} \frac{\big( \eps - \eps_b + (\eps + \eps_b) \exp{2 q (z_< + 2 d)} \big)  \big( \eps + \eps_a + (\eps - \eps_a) \exp{2 q (z_> - 2 d)} \big)}{(\eps_a+\eps)(\eps_b+\eps) - (\eps_a-\eps)(\eps_b-\eps) \exp{-8qd}},
\end{equation}
where $z_< = \text{min}(z,z')$ and $z_> = \text{max}(z,z')$. At this point, we can apply the ideal 2D-limit $z=d$, $z'=-d$ and calculate the interlayer potential as
\begin{equation}
    w(q, -d, d) = \frac{e^2}{2 q \eps \eps_0} \frac{\big( \eps - \eps_b + (\eps + \eps_b) \exp{2 q d} \big)  \big( \eps + \eps_a + (\eps - \eps_a) \exp{-2 q d} \big)}{(\eps_a+\eps)(\eps_b+\eps)  \exp{4qd} - (\eps_a-\eps)(\eps_b-\eps) \exp{-4qd}}.
\end{equation}
For the intralayer potential we have two cases, namely both carriers in the top layer $z = z' = d$ or bottom layer $z = z' = -d$. In the case of differing super- and substrates, these cases will be slightly different. However, assuming identical sub- and superstrates, $\eps_a = \eps_b$, the two cases will be equal due to symmetry, thus
\begin{equation}
    w(q, d, d) = w(q, -d, -d) = \frac{e^2 \exp{-2qd}}{2 q \eps \eps_0} \frac{\big( \eps - \eps_b + (\eps + \eps_b) \exp{6 q d} \big)  \big( \eps + \eps_a + (\eps - \eps_a) \exp{-2q d} \big)}{(\eps_a+\eps)(\eps_b+\eps) \exp{4qd} - (\eps_a-\eps)(\eps_b-\eps) \exp{-4qd}}.
\end{equation}
The dielectric constant can be approximated from the static 2D sheet susceptibility as $\eps = \frac{\pi}{d} \chi_0 + 1$, where the static 2D sheet susceptibility $\chi_0$ is calculated from the DFT band structure.

\subsection{Coulomb and exchange kernels in the bilayer 2D-limit}
\label{sec:framework}
In this appendix, we shall establish an expression for the Coulomb and exchange kernels in a bilayer material. We start from Bloch's theorem and write the lattice-periodic part of the wave function as $u_{n,\Vec{k}}(\Vec{r}) = \exp{-i \Vec{k} \cdot \Vec{r}} \psi_{n,\Vec{k}}(\Vec{r})$, which is then expanded in 2D Fourier coefficients using the periodicity of the unit cell
\begin{equation}
   u_{n,\Vec{k}}(\Vec{r}) = \frac{1}{\sqrt{A}} \sum_{\Vec{G}} C_{n,\Vec{k}}(\Vec{G}, z) \exp{i \Vec{G} \cdot \vec{\rho}},
\end{equation}
where $\Vec{\rho}$ is the in-plane coordinate and $A$ the area. Next, products of these are Fourier decomposed, such that
\begin{equation}
u_{n,\Vec{k}}^*(\Vec{r}) u_{m,\Vec{k}'}(\Vec{r}) = \frac{1}{A} \sum_{\Vec{G}} i_{n\Vec{k},m\Vec{k}'}(\Vec{G}, z) \exp{i \Vec{G} \cdot \vec{\rho}},
\end{equation}
where the Fourier coefficients are given as 
\begin{equation}
    i_{n\Vec{k},m\Vec{k}'}(\Vec{G}, z) = \int_{A} u_{n,\Vec{k}}^*(\Vec{r}) \exp{-i \Vec{G} \cdot \vec{\rho}} u_{m,\Vec{k}'}(\Vec{r}) d^2 \vec{\rho} = \sum_{\Vec{G}'} C_{n \Vec{k}}^* (\Vec{G}' - \Vec{G}, z) C_{m \Vec{k}'} (\Vec{G}', z).
\end{equation}
Furthermore, in the bilayer 2D-limit where an electron or hole will always be strictly located to either of the two layers, that is $z=\pm d$, we can write
\begin{equation}
    i_{n\Vec{k},m\Vec{k}'}(\Vec{G}, z) \simeq I^-_{n\Vec{k},m\Vec{k}'}(\Vec{G}) \delta(z+d) + I^+_{n\Vec{k},m\Vec{k}'}(\Vec{G}) \delta(z-d),
    \label{eq:bilayer}
\end{equation}
where
\begin{equation}
    \begin{split}
        I^-_{n\Vec{k},m\Vec{k}'}(\Vec{G}) &= \int_{-\infty}^0 i_{n\Vec{k},m\Vec{k}'}(\Vec{G}, z) dz, \\
        I^+_{n\Vec{k},m\Vec{k}'}(\Vec{G}) &= \int_{0}^\infty i_{n\Vec{k},m\Vec{k}'}(\Vec{G}, z) dz. \\
    \end{split}
\end{equation}
In a similar manner, the Coulomb interaction $\op{W}(\Vec{r},\Vec{r}')$ is Fourier decomposed in the entire 2D-plane
\begin{equation}
    \op{W}(\Vec{r},\Vec{r}') = \frac{1}{A} \sum_{\Vec{q},\Vec{G}} w(\Vec{q}+\Vec{G}, z, z') \exp{i(\Vec{q}+\Vec{G})\cdot(\vec{\rho}-\vec{\rho}')},
\end{equation}
where $\Vec{q}$ is a 2D vector confined to the Brillouin zone, and $w(\Vec{q}+\Vec{G}, z, z')$ is the Coulomb potential in 2D Fourier space. The elements of the Coulomb interaction are then
\begin{equation}
\begin{split}
    \braket{1,2|\op{W}(\Vec{r},\Vec{r}')|3,4} &= \frac{1}{A^3} \sum_{\Vec{q}, \Vec{G}, \vec{G}_{1,3},\vec{G}_{2,4}} \int_A \exp{i \left(\Vec{G} + \Vec{G}_{1,3} + \Vec{q} + \Vec{k}_3 - \Vec{k}_1 \right) \cdot \vec{\rho}} d^2 \vec{\rho} \int_A \exp{-i \left(\Vec{G} - \Vec{G}_{2,4} + \Vec{q} - \Vec{k}_4 + \Vec{k}_2 \right) \cdot \vec{\rho}} d^2 \vec{\rho}  \\  
    &\times \int_{-\infty}^{\infty} \int_{-\infty}^{\infty} i_{1,3}(\Vec{G}_{1,3}, z) i_{2,4}(\Vec{G}_{2,4}, z') w( \Vec{q}+\Vec{G}, z - z') d z d z' .
\end{split}
\end{equation}
Due to the integrals over area, the non-zero terms in the sums must have vanishing exponents in the integrand. Furthermore, since $\Vec{q}$ and $\Vec{k}$ are restricted to the Brillouin zone, we get $\Vec{q} = \Vec{k}_4 - \Vec{k}_2 = \Vec{k}_1 - \Vec{k}_3$ and $\Vec{G} = -\Vec{G}_{1,3} = \Vec{G}_{2,4}$. Now, the $z,z'$ integral can be rewritten using Eq. \eqref{eq:bilayer}
\begin{equation}
\begin{split}
    &\int_{-\infty}^{\infty} \int_{-\infty}^{\infty} i_{1,3}(-\Vec{G}, z) i_{2,4}(-\Vec{G}, z') w(\Vec{q}+\Vec{G}, z - z') d z d z' \\
    = \, & w(\Vec{q}+\Vec{G}, 0) I^{-*}_{3,1} (\Vec{G}) I^-_{2,4} (\Vec{G}) + w(\Vec{q}+\Vec{G}, 2d) I^{-*}_{3,1} (\Vec{G}) I^+_{2,4} (\Vec{G}) \\
    + \, & w(\Vec{q}+\Vec{G}, 2d) I^{+*}_{3,1} (\Vec{G}) I^-_{2,4} (\Vec{G}) + w(\Vec{q}+\Vec{G}, 0) I^{+*}_{3,1} (\Vec{G}) I^+_{2,4} (\Vec{G}) ,
\end{split}
\end{equation}
where Hermiticity $I_{1, 2}(\Vec{G}) = I_{2, 1}^*(-\Vec{G})$ has been applied. Here, $w(\Vec{q}+\Vec{G}, 0)$ is the intralayer potential and $w(\Vec{q}+\Vec{G}, 2d)$ the interlayer potential. The Coulomb ($C$) and exchange ($x$) kernels thus become
\begin{align}
\begin{split}
    &\braket{v' \Vec{k}' - \Vec{Q}/2,c \Vec{k} + \Vec{Q}/2|\op{W}_C(\Vec{r},\Vec{r}')|v \Vec{k} - \Vec{Q}/2,c' \Vec{k}' + \Vec{Q}/2} =  \frac{1}{A} \sum_{\Vec{G}} 
    \\ \times \Big[ & w_C(\Vec{k}' - \Vec{k} + \vec{G}, 0) \Big( I^{-*}_{v \Vec{k} - \Vec{Q}/2, v' \Vec{k}' - \Vec{Q}/2}(\Vec{G}) I^-_{c \Vec{k} + \Vec{Q}/2, c' \Vec{k}' + \Vec{Q}/2}(\Vec{G}) + I^{+*}_{v \Vec{k} - \Vec{Q}/2, v' \Vec{k}' - \Vec{Q}/2}(\Vec{G}) I^+_{c \Vec{k} + \Vec{Q}/2, c' \Vec{k}' + \Vec{Q}/2}(\Vec{G}) \Big)
    \\ + \, & w_C(\Vec{k}' - \Vec{k} + \vec{G}, 2d) \Big( I^{-*}_{v \Vec{k} - \Vec{Q}/2, v' \Vec{k}' - \Vec{Q}/2}(\Vec{G}) I^+_{c \Vec{k} + \Vec{Q}/2, c' \Vec{k}' + \Vec{Q}/2}(\Vec{G}) + I^{+*}_{v \Vec{k} - \Vec{Q}/2, v' \Vec{k}' - \Vec{Q}/2}(\Vec{G}) I^-_{c \Vec{k} + \Vec{Q}/2, c' \Vec{k}' + \Vec{Q}/2}(\Vec{G}) \Big) \Big],
\end{split} \label{eq:coulombPart}\\
\begin{split}
   &\braket{v' \Vec{k}' - \Vec{Q}/2,c \Vec{k} + \Vec{Q}/2|\op{W}_x(\Vec{r},\Vec{r}')|c' \Vec{k}' + \Vec{Q}/2, v \Vec{k} - \Vec{Q}/2} =  \frac{1}{A} \sum_{\Vec{G}} \\
   \times \Big[ &w_x(\vec{G} - \Vec{Q}, 0) \Big ( I^{-*}_{c' \Vec{k}' + \Vec{Q}/2, v' \Vec{k}' - \Vec{Q}/2}(\Vec{G}) I^-_{c \Vec{k} + \Vec{Q}/2, v \Vec{k} - \Vec{Q}/2}(\Vec{G}) + I^{+*}_{c' \Vec{k}' + \Vec{Q}/2, v' \Vec{k}' - \Vec{Q}/2}(\Vec{G}) I^+_{c \Vec{k} + \Vec{Q}/2, v \Vec{k} - \Vec{Q}/2}(\Vec{G})  \Big)
   \\ + \, & w_x(\vec{G} - \Vec{Q}, 2d) \Big( I^{+*}_{c' \Vec{k}' + \Vec{Q}/2, v' \Vec{k}' - \Vec{Q}/2}(\Vec{G}) I^-_{c \Vec{k} + \Vec{Q}/2, v \Vec{k} - \Vec{Q}/2}(\Vec{G}) + I^{-*}_{c' \Vec{k}' + \Vec{Q}/2, v' \Vec{k}' - \Vec{Q}/2}(\Vec{G}) I^+_{c \Vec{k} + \Vec{Q}/2, v \Vec{k} - \Vec{Q}/2}(\Vec{G}) \Big) \Big].
\end{split} \label{eq:exchangePart}
\end{align}

\subsection{Electron and hole densities}
\label{sec:wfs}
\begin{figure}[htp]
    \centering
    \includegraphics[trim={0cm 0.0cm 1.5cm 0.5cm},clip,width=.5\linewidth]{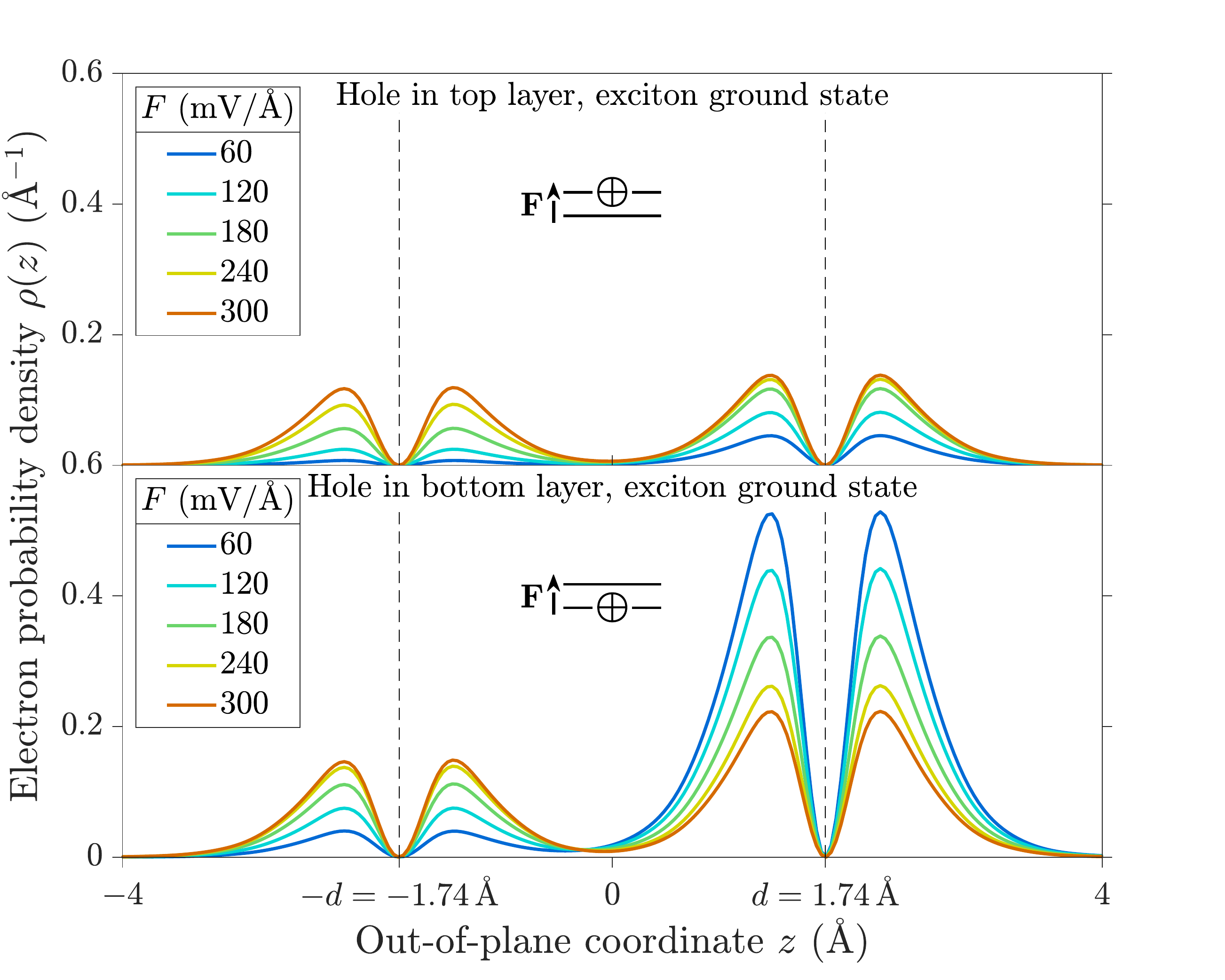}
    \caption{(a) In-plane averaged electron probability density $\rho(z)$ plotted as a function of the out-of-plane coordinate $z$ for a hole placed in either the lower or upper layer.}
    \label{fig:wfs}
\end{figure}

To obtain expressions for the electron and hole densities, we start with the exciton wave function of the $\lambda$ exciton, given as
\begin{equation}
    \Psi^\lambda(\vec{r}_e, \vec{r}_h) = \sum_{\vec{k} c v} A_{\vec{k} c v}^\lambda \psi_{c, \Vec{k}}(\vec{r}_e) \psi_{v, \Vec{k}}(\vec{r}_h),
    \label{eq:excitonWFS}
\end{equation}
where $\psi_{n, \Vec{k}}(\vec{r}) = u_{n, \Vec{k}}(\vec{r}) \exp{i \Vec{k} \cdot \vec{r}}$ and $A_{\vec{k} c v}^\lambda$ the eigenvector of the BSE eigenvalue problem for $\vec{Q}=0$. In particular, we are interested in the two-particle density $\rho^{\lambda} = \abs{\Psi^\lambda}^2$, thus
\begin{equation}
    \rho^\lambda(\vec{r}_e, \vec{r}_h) = \sum_{\vec{k}' c' v'} \sum_{\vec{k} c v} A_{\vec{k}' c' v'}^{\lambda *} A_{\vec{k} c v}^\lambda \psi_{c, \Vec{k}}(\vec{r}_e) \psi^*_{c', \Vec{k}'}(\vec{r}_e) \psi_{v, \Vec{k}}(\vec{r}_h) \psi^*_{v', \Vec{k}'}(\vec{r}_h).
\end{equation}
Since the density is a function of both coordinates, and we are only interested in determining the out-of-plane dependence of one of them, we shall do so by fixing the hole to either the top or bottom layer. That is, we shall average the density over said layer, thus, in the case of a hole in the top layer
\begin{align}
    \rho_e^\lambda(\vec{r}_e, z_h>0) &= \int_{z_h > 0} \rho^\lambda(\vec{r}_e, \vec{r}_h) \, d^3 \vec{r}_h \\
    &= \sum_{\vec{k} c v} \sum_{\vec{k}' c' v'} A_{\vec{k}' c' v'}^{\lambda *} A_{\vec{k} c v}^\lambda \psi_{c, \Vec{k}}(\vec{r}_e) \psi^*_{c', \Vec{k}'}(\vec{r}_e) \int_{z_h > 0} \psi_{v, \Vec{k}}(\vec{r}_h) \psi^*_{v', \Vec{k}'}(\vec{r}_h) \, d^3 \vec{r}_h.
\end{align}
Now, due to in-plane orthogonality, only terms having $\vec{k} = \vec{k}'$ survive, and we get
\begin{equation}
    \rho_e^\lambda(\vec{r}_e, z_h>0) = \sum_{\vec{k} c v} \sum_{c' v'} A_{\vec{k} c' v'}^{\lambda *} A_{\vec{k} c v}^\lambda \psi_{c, \Vec{k}}(\vec{r}_e) \psi^*_{c', \Vec{k}}(\vec{r}_e) \int_{z_h > 0} \psi_{v, \Vec{k}}(\vec{r}_h) \psi^*_{v', \Vec{k}}(\vec{r}_h) \, d^3 \vec{r}_h.
\end{equation}
Next, we average the in-plane coordinates of the electron to obtain the $z$-dependence
\begin{align}
    \rho_e^\lambda(z, z_h>0) &= \int \rho_e^\lambda(\vec{r}_e) \, d x \, d y \\
    &= \sum_{\vec{k} c v} \sum_{c' v'} A_{\vec{k} c' v'}^{\lambda *} A_{\vec{k} c v}^\lambda \int \psi_{c, \Vec{k}}(\vec{r}_e) \psi^*_{c', \Vec{k}}(\vec{r}_e) \, d x \, d y \, \int_{z_h > 0} \psi_{v, \Vec{k}}(\vec{r}_h) \psi^*_{v', \Vec{k}}(\vec{r}_h) \, d^3 \vec{r}_h,
\end{align}
and similarly for a hole in the bottom layer
\begin{equation}
    \rho_e^\lambda(z, z_h<0) = \sum_{\vec{k} c v} \sum_{c' v'} A_{\vec{k} c' v'}^{\lambda *} A_{\vec{k} c v}^\lambda \int \psi_{c, \Vec{k}}(\vec{r}_e) \psi^*_{c', \Vec{k}}(\vec{r}_e) \, d x \, d y \, \int_{z_h < 0} \psi_{v, \Vec{k}}(\vec{r}_h) \psi^*_{v', \Vec{k}}(\vec{r}_h) \, d^3 \vec{r}_h.
\end{equation}
Plotting these electron probability densities yields Fig. \ref{fig:wfs}. Finally, to obtain the probabilities shown in Fig. \ref{fig:bandCross}c, one needs only integrate over the electron coordinate in these expressions within the relevant layers. 

\twocolumngrid

\bibliography{bibliography} 

\end{document}